%% file: paper.tex
\documentclass[sigconf]{acmart}
\AtBeginDocument{%
  \providecommand\BibTeX{{%
    Bib\TeX}}}

\setcopyright{none}

\usepackage{amsmath,amsfonts}
\usepackage{graphicx}
\usepackage{textcomp}
\usepackage{xcolor}
\usepackage{listings}
\usepackage{xspace}
\usepackage{subfig}
\usepackage{soul}
\usepackage[export]{adjustbox}
\usepackage{algorithm}
\usepackage{algpseudocode}
\usepackage{multicol}

\newcommand{\tweakedsim}{\raise.17ex\hbox{$\scriptstyle\mathtt{\sim}$}}

\definecolor{MyRed}{HTML}{CF0000}
\definecolor{MyGreen}{HTML}{009E73}
\newcommand{\grcheck}{{\color{MyGreen}\checkmark}}
\newcommand{\rdcross}{{\color{MyRed}$\times$}}

\definecolor{dkgreen}{RGB}{0,64,0}
\definecolor{ltgray}{RGB}{245,245,245}
\definecolor{mauve}{RGB}{139,0,139}

\def\BibTeX{{\rm B\kern-.05em{\sc i\kern-.025em b}\kern-.08em
    T\kern-.1667em\lower.7ex\hbox{E}\kern-.125emX}}

\begin{document}

\title{Automated Programmatic Performance Analysis of Parallel Programs}

\author{Onur Cankur}
\email{ocankur@umd.edu}
\affiliation{%
    \institution{Department of Computer Science, University of Maryland}
    \city{College Park}
    \state{Maryland}
    \country{USA}
    \postcode{20742}
}

\author{Aditya Tomar}
\email{adityatomar@berkeley.edu}
\affiliation{%
    \institution{Department of Electrical Engineering and Computer Sciences, University of California, Berkeley}
    \city{Berkeley}
    \state{California}
    \country{USA}
}

\author{Daniel Nichols}
\email{dnicho@umd.edu}
\affiliation{%
    \institution{Department of Computer Science, University of Maryland}
    \city{College Park}
    \state{Maryland}
    \country{USA}
    \postcode{20742}
}

\author{Connor Scully-Allison}
\email{cscullyallison@email.arizona.edu}
\affiliation{%
    \institution{Department of Computer Science, The University of Arizona}
    \state{Arizone}
    \country{USA}
}

\author{Katherine E. Isaacs}
\email{kisaacs@sci.utah.edu}
\affiliation{%
    \institution{Department of Computer Science, The University of Utah}
    \state{Utah}
    \country{USA}
}

\author{Abhinav Bhatele}
\email{bhatele@cs.umd.edu}
\affiliation{%
    \institution{Department of Computer Science, University of Maryland}
    \city{College Park}
    \state{Maryland}
    \country{USA}
    \postcode{20742}
}

\renewcommand{\shortauthors}{Bhatele et al.}

\begin{abstract}
    \input{abstract}
\end{abstract}


\maketitle

\section{Introduction}
\label{sec:intro}
\input{intro}

\section{Background and Related Work}
\label{sec:background}
\input{background}

\section{Simplifying Performance Analysis Tasks}
\label{sec:approach}
\input{approach}

\section{Chopper: A Python API for Performance Analysis}
\label{sec:api}
\input{api}

\section{Experimental Setup}
\label{sec:setup}
\input{setup}

\section{Performance Evaluation of Chopper}
\label{sec:performance}
\input{performance}

\section{Case Studies}
\label{sec:case_study}
\input{case_study}

\section{Conclusion}
\label{sec:conclusion}
\input{conclusion}

\bibliographystyle{ACM-Reference-Format}
\bibliography{./bib/cite,./bib/pssg}

\end{document}

%% file: abstract.tex
Developing efficient parallel applications is critical to advancing scientific
development but requires significant performance analysis and optimization.
Performance analysis tools help developers manage the increasing complexity
and scale of performance data, but often rely on the user to manually explore
low-level data and are rigid in how the data can be manipulated. We propose a
Python-based API, Chopper, which provides high-level and flexible performance
analysis for both single and multiple executions of parallel applications.
Chopper facilitates performance analysis and reduces developer effort by
providing configurable high-level methods for common performance analysis
tasks such as calculating load imbalance, hot paths, scalability bottlenecks,
correlation between metrics and CCT nodes, and causes of performance variability
within a robust and mature Python environment that provides fluid access to
lower-level data manipulations. We demonstrate how Chopper allows developers
to quickly and succinctly explore performance and identify issues across
applications such as AMG, Laghos, LULESH, Quicksilver and Tortuga.

%% file: intro.tex
Ensuring that parallel applications run efficiently on modern supercomputers
is essential to achieve scientific discoveries rapidly.  Identifying
performance problems is the first step in the process of optimizing
performance of a parallel program. However, performance analysis is a complex
and time-consuming task due to the inherent complexity of large-scale parallel
applications and architectures, and the large quantity of performance data
that can be collected when running in parallel.  In addition, parallel
applications may suffer from a variety of performance issues.  Therefore,
in order to minimize the developer's burden, we require highly effective
performance analysis techniques that can quickly identify performance problems
and their root causes.

A variety of performance measurement tools exist, including profilers and
tracing tools that can generate performance data~\cite{caliper,hpctoolkit,scorep,shende:tau2006}.
However, the data generated can be extremely large, making it challenging to
sift through this data to identify performance issues.  Several performance
measurement tools also provide visual analytics counterparts to facilitate
performance analysis.  Typically, the analysis support is in the form of a
graphical user interface (GUI) to visualize and manipulate performance
data~\cite{mellorcrummey+:jsc02,huck:perfexplorer2005,bell:paraprof2003, knupfer2008vampir}
although some tools also provide a scripting
interface~\cite{saviankou2015cube, extrap}.  The GUIs help in visualizing
performance data and in many cases, the user can connect such data to source
code (file and line numbers).

Although GUIs provide some effective functionalities, having to analyze
performance data only via a GUI can make it inefficient to identify
performance issues. GUIs depend on the end user to manually explore
the visualizations, and to identify different patterns that might suggest
performance problems. As the data being analyzed grows, this becomes
more and more challenging. In addition, to analyze multiple executions,
GUIs typically require opening multiple separate windows with the datasets.
Some of them can visualize multiple datasets on the same window, however,
they still require significant manual effort to compare different executions.
Finally, adding new kinds of analyses on top of a GUI may not be
possible for an end user.

The main aim of this work is to simplify common performance analysis tasks
for the end user by reducing the time and effort required.
We present a Python-based API that simplifies several performance analysis
tasks, and offers flexibility and customization to enable users to
perform analyses with speed and effectiveness. To achieve this goal,
we explored the common functionalities in other performance analysis tools
and also collected feedback from developers
and users of performance tools to identify the most needed functionalities.
We develop this new API on top of an existing open-source performance analysis
tool called Hatchet that provides an interface for programmatic analysis of
performance data via Python ~\cite{bhatele:sc2019}.

By virtue of being developed on top of Hatchet, Chopper supports data
formats of various performance tools, including but not limited to
Caliper~\cite{caliper}, HPCToolkit~\cite{hpctoolkit},
Score-P (Cubex)~\cite{scorep} and TAU~\cite{shende:tau2006}. Chopper facilitates
performance analysis and reduces developer effort by simplifying tasks such as
detection of load imbalance, hot paths, scalability bottlenecks, and causes
of performance variability via a programmatic interface.  It provides
support for analyzing profiles from single and multiple executions of programs.
By using the provided functionalities in Chopper, users can quickly and
easily identify performance issues in a parallel program with a few
lines of Python code.  Since it is a programmatic API, Chopper gives
flexibility to the users to extend it and also use other Python libraries
with it for visualization and further analysis. To demonstrate the
usability and flexibility of Chopper, we gather performance data using
several applications including the data collected in a prior performance
variability study~\cite{nichols:ipdps2022}. The applications
used in this study are AMG, Laghos, LULESH, Quicksilver and Tortuga.

Specifically, this work makes the following contributions:
\begin{itemize}
      \item A programmatic API that significantly simplifies
            several single-run performance analysis tasks.
      \item Facilitate the analysis of multiple executions by designing and
            implementing algorithms for multi-run analysis that enable an
            effective and intuitive approach to identifying performance issues
            across multiple executions.
      \item An evaluation of the scalability of some user-facing functions provided in
            Chopper by using large parallel profiles.
      \item Demonstration of the usefulness of Chopper and its capabilities to
            identify performance issues by performing case studies using
            several applications.
\end{itemize}

%% file: background.tex
We give background information on profiling, call graphs,
and common performance analysis techniques. We also mention
Hatchet and other related work.
\begin{table*}[t]
    \caption{Capabilities in different profile analysis and visualization tools.}
    \label{tab:survey}
    \begin{tabular}{lccccccccccc} \toprule
                     & Hot Path & Load Imb. & Programmatic & Flat     & Speedup  & Correlation & Multirun & Perf.    \\
                     & Analysis & Analysis  & API          & Profile  & Analysis & Analysis    & Analysis & Modeling \\ \midrule
        Cube         & Manual   & \rdcross  & \rdcross     & Guided   & Manual   & \rdcross    & Manual   & \rdcross \\
        Extra-P      & \rdcross & \rdcross  & \rdcross     & \rdcross & Guided   & \grcheck    & Guided   & \grcheck \\
        Hatchet      & Manual   & Manual    & \grcheck     & Manual   & Manual   & \rdcross    & Manual   & \rdcross \\
        hpcviewer    & Guided   & Manual    & \rdcross     & Guided   & Manual   & \rdcross    & Manual   & \rdcross \\
        ParaProf     & Manual   & Manual    & \rdcross     & Guided   & Manual   & \rdcross    & Manual   & \rdcross \\
        PerfExplorer & \rdcross & \rdcross  & \rdcross     & \rdcross & Guided   & \grcheck    & Guided   & \rdcross \\
        Thicket      & Manual   & Manual    & \grcheck     & Manual   & Guided   & \rdcross    & Guided   & \grcheck \\
        This work    & Guided   & Guided    & \grcheck     & Guided   & Guided   & \grcheck    & Guided   & \rdcross \\ \bottomrule
    \end{tabular}
\end{table*}

\subsection{Profiling and Call Graphs}
Generally, there are two methods for performance measurement:
profiling and tracing. Profiling provides a statistical
approximation instead of exact
timestamps for each event in the program, unlike
tracing. In this paper we focus on profiling.

The performance data generated by profiling tools
provide a variety of information such as function
call sequences, performance metrics (e.g., time,
cache misses, floating-point operations per second),
and MPI process topologies. In this work we
study the analysis of calling context trees (CCT)
and call graphs, in both of which the nodes typically
represent procedures and edges represent the caller-callee
relationships (i.e., function call sequences). A path
from any given node to the root is called a call path
or calling context. A collection of
distinct calling contexts forms a CCT. Unlike CCTs, in
call graphs a procedure that is called in different
call paths is represented as a single node with aggregated
metric values. Therefore, call graphs are less
context-sensitive, but they provide a more concise
representation. Additionally, performance data typically
contains information about function names, file names, line
numbers, and process or thread IDs.

\subsection{Hatchet}

Hatchet is a Python-based tool that enables
analyzing hierarchical data, such as CCTs and
call graphs, programmatically~\cite{bhatele:sc2019}.
It reads performance data from several profiling tools
(e.g., Caliper, HPCToolkit, Score-P, TAU, timemory)
and provides an interface for programmatic analysis of
performance data. It also provides several visualization
functionalities such as terminal, DOT, and interactive
Jupyter notebook visualization.
It supports low level operations to manipulate the data
and requires significant programming.

Hatchet's central data structure is called GraphFrame,
which is a combination of a pandas
DataFrame~\cite{mckinney:pandas,mckinney:pandas2}
and a Graph. It stores the caller-callee relationships
in the graph object and the associated performance metrics and
contextual information in the DataFrame. Hatchet provides
graph-indexed DataFrames, which means every index of the
DataFrame points to a node in the graph. Therefore, these
two data structures are connected and can be manipulated
together. This data structure enables the practical
implementation of different analysis tasks. We
utilize Hatchet to implement our analysis
API, Chopper.

\subsection{Common Performance Analysis Problems}
The performance of a parallel application can suffer
from communication or computational inefficiencies.
Performance problems can be revealed by investigating
imbalances, scalability, variability, and hot paths
in the program.

\vspace{0.08in}
\noindent{\textbf{Load Imbalance:}} Parallel
programs use multiple processing elements (e.g.,
processes and threads). Ideally, the work done
by the program should be equally distributed over
processing elements, so that they can finish
their tasks at the same time. However, the ideal
scenario is almost never achieved in complex
workflows, which makes the load imbalance a
common problem. It can be identified by investigating
the cost incurred by different processes.

\vspace{0.08in}
\noindent{\textbf{Hot Paths:}} One way to
pinpoint the bottlenecks in the program is
to examine the most time-consuming call paths. This
task is called hot path analysis~\cite{adhianto2010effectively}.
For a given metric (e.g., time), every node in
a hot path accounts for more than 50\% percent
of its parents. Manually finding the hot path
is a tedious task when there are millions of nodes
in the call graph.

\vspace{0.08in}
\noindent{\textbf{Poor Scalability:}} Scalability
analysis shows how well a program utilizes the increasing
number of processing elements. A program that has poor
scalability may work slower than expected despite using
more processing elements. Scalability problems can be
identified by performing scaling experiments and
observing the change in speedup and efficiency.

\vspace{0.08in}
\noindent{\textbf{Performance Variation:}} The performance
of a program may differ in different runs even though
all of the parameters used in each run are the same
(e.g., hardware architecture and input parameters). For example
network congestion can lead to variability in
performance~\cite{bhatele:sc2013}. Variability can be
identified by analyzing multiple identical runs of a program.

\subsection{Related Work}
The idea of analyzing single and multiple call graph
data to pinpoint bottlenecks is defined as differential
profiling by early work~\cite{mckenney:differential1995}.
Later works demonstrated the usefulness of manipulating
and visualizing call graph data by performing differential
analysis~\cite{schulz+:differential07}. With that knowledge
many studies utilized the call graph data to effectively
identify and visualize performance problems. Several studies
manipulated performance metrics to identify load
imbalances~\cite{tallent+:sc10,derose2007detectingload,huck2010detailedload}.
Adhianto et al.~\cite{adhianto2010effectively} defined
hot path analysis and demonstrated how to perform it
using HPCViewer. Several studies demonstrated applying
differential profiling by using call path profiles to
analyze the scalability of the programs ~\cite{tallent2009diagnosing,coarfa2007scalability, liu2015scaanalyzer}.
Benedict et al.~\cite{benedict2009automatic} examined
the scalability of the programs by instrumenting the
region of interests in the programs and analyzing the
performance of different processes on those regions.
Variability in performance of HPC applications is another
commonly studied topic ~\cite{PeKePa03}
However, to the best of our knowledge, there is no
study on analyzing performance variability using
call graph data.

\subsubsection{Performance Analysis Tools}
Many performance analysis tools are developed
to facilitate performance analysis. Table \ref{tab:survey}
provides a summary of different tools.
Cube~\cite{saviankou2015cube} is a performance analysis
tool for Score-P.
Extra-P~\cite{extrap} is an automated performance modeling tool
that focuses mostly on scaling behavior of applications.
HPCViewer~\cite{mellorcrummey+:jsc02}
enables analyzing profile and trace data generated by HPCToolkit.
ParaProf~\cite{bell:paraprof2003} is also a performance
analysis tool and a part of TAU toolkit. It supports several
different profile data formats. All of these tools present
call graph along with performance metrics. Additionally, Cube
and ParaProf can visualize MPI process topologies.
PerfExplorer~\cite{huck:perfexplorer2005} framework is also
a part of TAU toolkit and supports several data mining operations
such as correlation analysis and clustering.
Thicket~\cite{thicket}, provides performance analysis
capabilities for multi-run performance experiments. It
utilizes Hatchet and Extra-P and develop new capabilities on top of them.
Even though all of these tools provide useful analysis
capabilities, most of them provide only a desktop GUI.
However, GUIs typically are not as flexible and dynamic
as a programmatic interface and they do not provide rich APIs
to manipulate the profile data. This limitation becomes more
obvious when the data is very large and complex. Additionally,
making changes or adding new capabilities to a GUI is hardly
possible for the end user. Thicket provides programmatic analysis
capabilities but only focuses on multi-run analysis.

We propose a Python-based API, Chopper, to overcome these
limitations. Chopper facilitates performance analysis by
simplifying several single-run and multi-run performance
analysis tasks and making them easier and more intuitive to
perform. We utilize Hatchet's programmatic
interface and visualization capabilities to implement the analysis
tasks. With the Chopper API the users can identify
performance problems in their parallel programs by writing only
a few lines of Python code. Chopper reduces the effort and
time spent on performance analysis.

%% file: approach.tex
Performance analysis of parallel applications is complex, tedious,
time-consuming and challenging. This is partly due to the fact that a
significant burden of performance analysis falls on the end user (code
developer, performance engineer, etc.). Our primary goal in this work is to
simplify different performance analysis workflows as much as possible, and
to make it straightforward for the end user to identify common performance
issues in their parallel code.

We started with creating a glossary of different performance analysis tasks
that end users perform when dealing with parallel code. We asked end users
of Hatchet for recommendations of analysis tasks they would want to see in
a performance analysis library. We also analyzed the tasks
supported by the GUIs of performance analysis tools and identified the user
effort in using them to arrive at performance issues. We also identified
the gaps in current tools that could be supported in a
programmatic API.

We observed that when end users are conducting performance analysis, there
are two broad categories of tasks that are significantly different from each
other. One type of tasks involves analyzing the performance of a single
execution (a specific application running a specific input problem on a
specific architecture). This is often done when the user knows that there is
some performance issue or has recently modified the code and wants to
understand its performance impacts. The second type of tasks involve
analyzing multiple executions. Such analyses are done in a variety of
contexts -- studying scaling performance with increasing number of
processes/threads, comparing the impact of different inputs on performance, understanding performance variability across multiple executions,
etc. We will refer to these two types of tasks as single-run and multi-run
analysis respectively in the rest of the paper. We then classified all the
performance analysis tasks in the glossary we created into either single-run
or multi-run type of tasks.

The design and implementation of single-run and multi-run analysis tasks can
be significantly different. Hence, when designing Chopper, we considered
them and the design issues surrounding them separately. Chopper provides a
unified interface for invoking functions from either category. However,
single-run analysis tasks can also be invoked from a GraphFrame
object in Hatchet.

%% file: api.tex
We describe the design of API we implemented, called ``Chopper'' because it
helps manipulate calling context trees (and call graphs). Chopper facilitates
a range of analysis tasks for both single (\autoref{sec:single})
and multiple executions (\autoref{sec:multi}).

\subsection{Analyzing a Single Execution}
\label{sec:single}

Through Chopper, we add higher-level performance analysis operations to the
lower-level performance metrics and manipulations offered by Hatchet. To
provide a seamless experience, we augment the Hatchet GraphFrame (pandas
dataframe + graph) object so Chopper methods can be called directly. These
methods are also available directly through the Chopper API by passing the
GraphFrame object. We describe Chopper's single-run capabilities below.

\begin{figure}[ht]
    \centering
    \includegraphics[width=1\columnwidth]{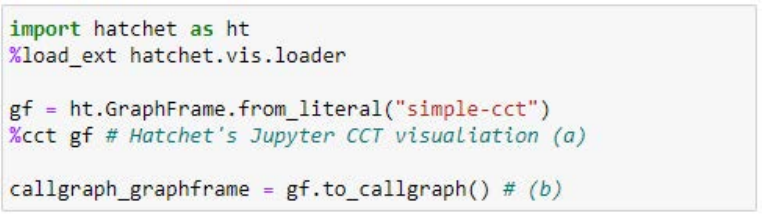}
    \newline
    \subfloat[CCT]{\includegraphics[width=0.94\columnwidth]{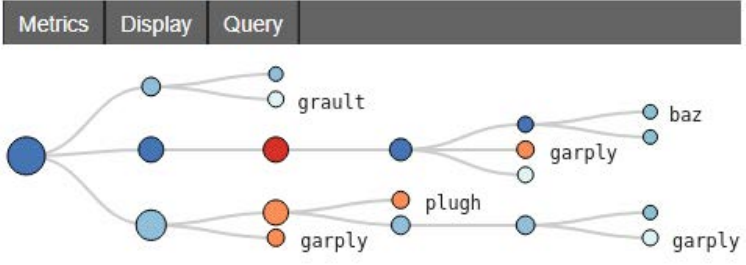}}
    \newline
    \subfloat[Call Graph]{\includegraphics[width=0.8\columnwidth]{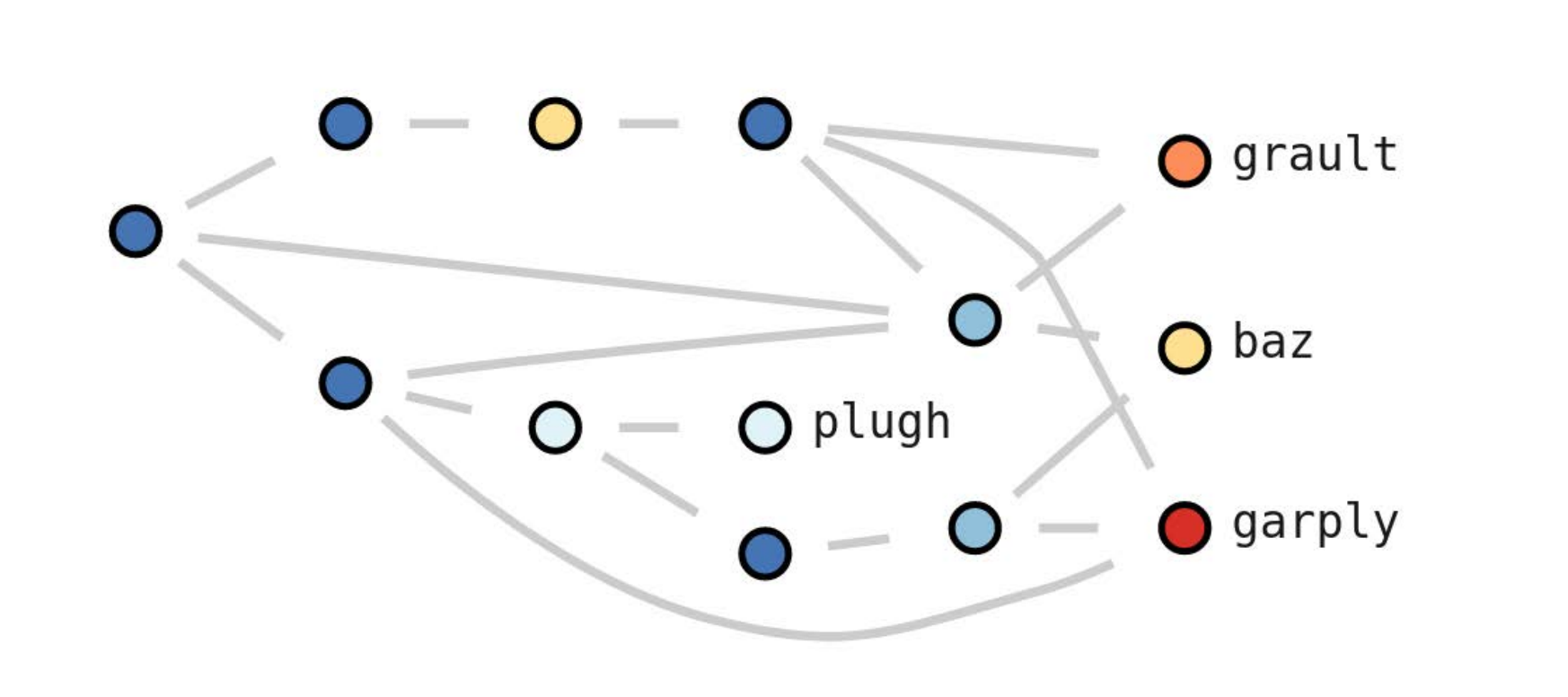}}
    \caption{Creating a callgraph from a CCT using the {\tt to\_callgraph} function.
        Hatchet's Jupyter
        notebook visualization is used to visualize the CCT (a). The
        call graph (b) is visualized externally.}
    \label{fig:cct_cg_trivial}
\end{figure}

\vspace{0.08in}
\noindent{\textbf{to\_callgraph}}: For some analyses, the full calling context
of each function is not necessary. It may be more intuitive to examine the
    {\em call graph} which merges all calls to the same function name into a single
node. The {\tt to\_callgraph} function converts a CCT into a call graph by
merging nodes representing the same function name and summing their associated
metric data. The output is a new GraphFrame where the graph has updated
(merged) caller-callee relationships and the DataFrame has aggregated metric
values.

Figure~\ref{fig:cct_cg_trivial} shows a small CCT and the resulting output
when converted to a
call graph using the to\_callgraph function. The call
graph representation is more concise but performance metrics are no longer
divided by calling context.

\begin{figure}[h]
    \centering
    \includegraphics[width=1\columnwidth]{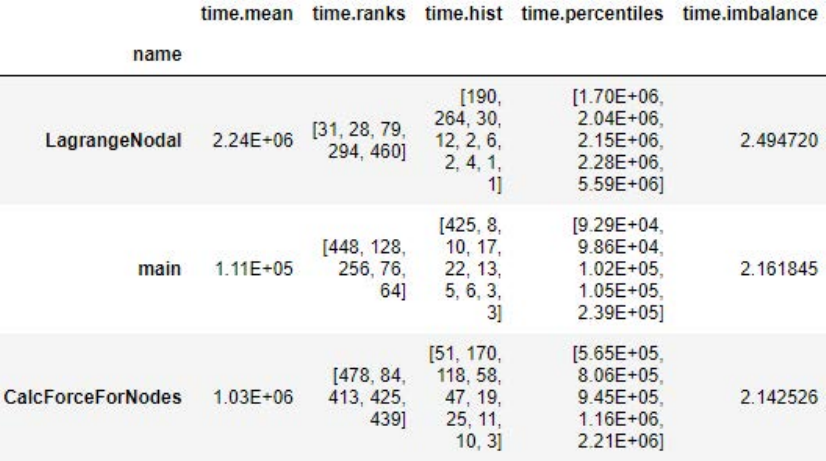}
    \includegraphics[width=1\columnwidth]{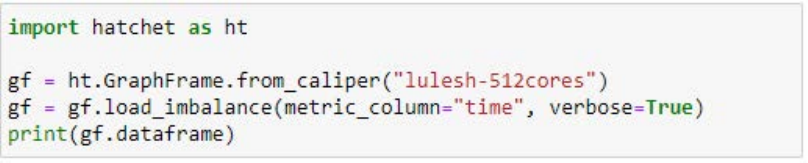}
    \includegraphics[width=1\columnwidth]{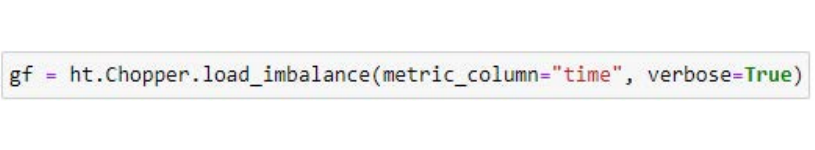}
    \caption{Calculating the load imbalance of a 512 process execution for
        LULESH by using the {\tt load\_imbalance} function. The resulting
        DataFrame is sorted by the time.imbalance column which shows the imbalance
        value for each CCT node.
    }
    \label{fig:load_imbalance_toy}
\end{figure}

\vspace{0.08in} \noindent{\textbf{load\_imbalance}}: Load imbalance is a
common performance problem in parallel programs. Developers and application
users are interested in identifying load imbalance so they can improve the
distribute of work among processes or threads. The {\tt load\_imbalance}
function in Chopper makes it easier to study load imbalance at the level of
individual CCT nodes.

Algorithm~\ref{alg:load_imb} summarizes the load\_imbalance function.
The input is a GraphFrame along with the metric on which to compute imbalance.
Optional parameters are a threshold value to filter out inconsequential nodes
and a flag for calculate detailed statistics about the load imbalance. The
output is a new GraphFrame with the same graph object but additional columns
in its DataFrame to describe load imbalance and optionally the verbose
statistics. A full example of load\_imbalance is shown in
Figure~\ref{fig:load_imbalance_toy}.

To calculate per-node load imbalance, we use pandas DataFrame operations to
compute the mean and maximum of the given metric across all processes (line 4
and 5). Load balance is then the maximum divided by the mean (line 15). A
large maximum-to-mean ratio indicates heavy load imbalance. The per-node load
imbalance value is added as a new column in DataFrame.

The threshold parameter is used to filter out nodes with metric values below
the given threshold (line 13). This feature allows users to remove nodes that
might have high imbalance because their metric values are small. For example,
high load imbalance may not have significant impact on overall performance in
the time spent in the node is small.

The verbose option calculates additional statistics. If
enabled, the function adds a new column to the resulting DataFrame with each
of the following: the top five ranks that have the highest metric values (line
7), values of 0th, 25th, 50th, 75th, and 100th percentiles of each node (line
8), and the number of processes in each of ten equal-sized bins between the
0th (minimum across processes) and 100th (maximum across processes) percentile
values (line 9).

\begin{algorithm}[h]
    \caption{Pseudocode of load\_imbalance} \label{alg:load_imb}
    {\small
        \begin{algorithmic}[1]
            \Function{load\_imbalance}{graphframe, metric, threshold, verbose} \label{func:load_imb}
            \State dataframe $\gets$ graphframe.dataframe
            \For {nodes $\in$ dataframe}
            \State dataframe[``metric.max"] $\gets$ max across processes
            \State dataframe[``metric.mean"] $\gets$ mean across processes
            \If {verbose}
            \State dataframe[``metric.ranks"] $\gets$ top five ranks
            \State dataframe[``metric.percentile"] $\gets$ percentile values
            \State dataframe[``metric.hist"] $\gets$ frequency histogram
            \EndIf
            \EndFor

            \If {threshold}
            \State dataframe $\gets$ filter(dataframe[``metric.max"] $>$ threshold)
            \EndIf
            \State dataframe[``metric.imbalance"] $\gets$ dataframe[``metric.max"] / dataframe[``metric.mean"]
            \State \Return {graphframe}
            \EndFunction
        \end{algorithmic}
    }
\end{algorithm}

\begin{figure}[ht]
    \centering
    \includegraphics[width=1\columnwidth]{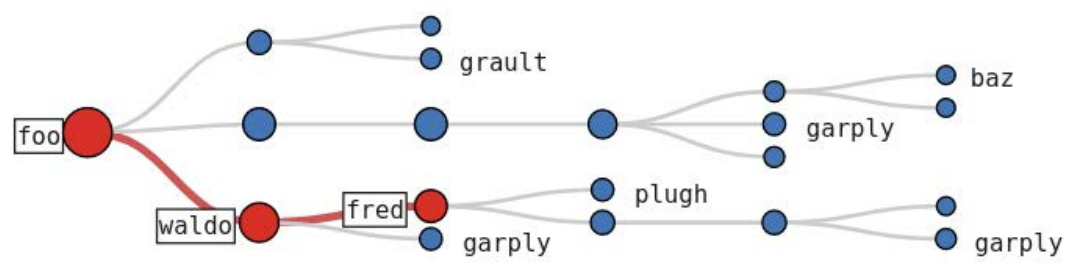}
\begin{lstlisting}[linewidth=1\columnwidth]
import hatchet as ht
%load_ext hatchet.vis.loader

gf = ht.GraphFrame.from_hpctoolkit("simple-profile")
hot_path = gf.hot_path()
%cct gf #Jupyter CCT visualiation
\end{lstlisting}
    \caption{Identifying the hot path of a simple CCT using the
            {\tt hot\_path} function in Chopper. The red-colored path with bigger,
        labeled nodes represents the hot path.}
    \label{fig:hot_path_toy}
\end{figure}

\vspace{0.08in}
\noindent{\textbf{hot\_path}}: A common task in analyzing a single execution
is to examine the most time-consuming call paths in the program or some subset
of the program. Seeking out the latter call paths can be tedious in a GUI,
especially if the CCT is large and complex. Chopper's {\tt hot\_path}
function retrieves the hot path from any subtree of a CCT given its root. The
input parameters are the GraphFrame, metric (and optional stopping condition),
and the root of the subtree to search. Starting at the given subtree root, the
method traverses the graph it finds a node whose metric accounts for more than a
given percentage of that of its parent. This percentage is the stopping
condition.
The hot path is then the path between that node and the given subtree
root. The function outputs a list of nodes using which the DataFrame can
be manipulated.

By default, the hot\_path function uses the most time-consuming root
node (in case of a forest) as the subtree root. The default stopping condition
is 50\%, which we chose based on its utility as identified by Adhianto et
al.~\cite{adhianto2010effectively}. The resulting hot path can be visualized
in the context of the CCT using the interactive Jupyter visualization in
Hatchet. We validated our implementation by comparing our results with
hpcviewer.

Figure~\ref{fig:hot_path_toy} shows the hot path for a simple CCT example,
found with a single Chopper function call (line 5) and visualized using
Hatchet's Jupyter notebook visualization (line 6).
The red-colored path with the large red nodes and additional labeling
represents the hot path. Users can interactively expand or collapse subtrees
to investigate the CCT further.

\vspace{0.08in}
\noindent{\textbf{correlation\_analysis}}:
Profiling data may include numerous metrics and CCT nodes and it is important to
analyze correlation between them to understand the program behaviour. To facilitate
this analysis, the Chopper API provides two main functions: {\tt correlation\_analysis}
and {\tt pairwise\_correlation}. The correlation\_analysis function calculates
the correlation between different performance metrics such as time, cache misses,
and branch misses. It accepts a GraphFrame, list of metrics, and a method to
calculate correlations (e.g., Pearson, Spearman, Kendall). It outputs the
correlation matrix. In order to simplify the analysis, Chopper provides
the {\tt filter\_correlation\_matrix} function that filters the
correlation matrix based on the correlation value. The pairwise\_correlation function
provides a more granular view, examining the relationship of two metrics at the
level of individual CCT nodes. This function performs linear regression and fits a
linear model to the data, assuming linear relationship between two performance
metrics. The CCT nodes that diverge significantly from the fitted line might imply
unusual behavior within the program and aid users to identify potential issues.
The pairwise\_correlation function adds the values on the regression line and the
distances of each CCT node to the GraphFrame's DataFrame.

\begin{figure}[t]
    \centering
    \includegraphics[width=1\columnwidth]{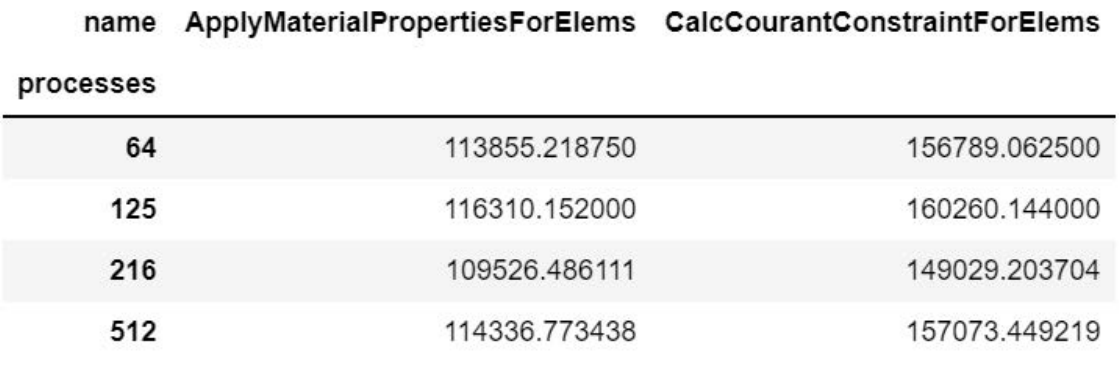}
\begin{lstlisting}[linewidth=1\columnwidth]
datasets = glob.glob("list_of_lulesh_profiles")
gfs = hatchet.GraphFrame.construct_from(datasets)
table = hatchet.Chopper.multirun_analysis(gfs)
print(table)
\end{lstlisting}
    \caption{The {\tt multirun\_analysis} function returns a pivot table
        containing node names and time values of the nodes in each profile. We
        show a truncated example of the returned pivot table from a set of
        LULESH weak scaling executions (64, 125, 216, and 512 processes).}
    \label{fig:construct-multirun}
\end{figure}

\begin{figure}[h]
    \centering
    \includegraphics[width=1\columnwidth]{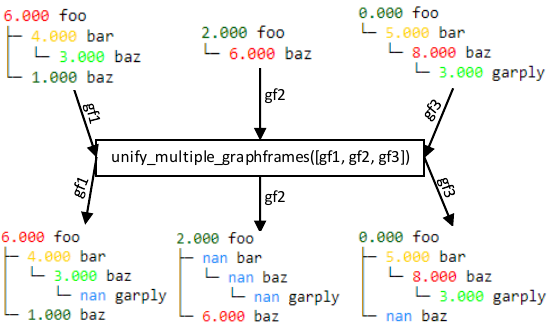}
    \caption{GraphFrames before and after unification by the
            {\tt unify\_multiple\_graphframes} function. The resulting GraphFrames include all nodes from the given GraphFrames
        but retain their original metric values.
    }
    \label{fig:unify-multiple-graphframes}
\end{figure}

\subsection{Comparing Multiple Executions}
\label{sec:multi}

Performance often only makes sense in the context of multiple executions, for
example, understanding weak or strong scaling. However, GUI-based performance
tools are often focused on single executions. While Hatchet has a few simple
pairwise operations on two GraphFrames (i.e., two executions), three or more
executions require programming ad hoc analyses. Chopper implements several
capabilities for comparing performance across several executions, targeting
common analyses such as those in studies of scaling scaling or performance
variability. These are implemented as static functions of the Chopper API.

\begin{figure}[h]
    \centering
    \includegraphics[width=1\columnwidth]{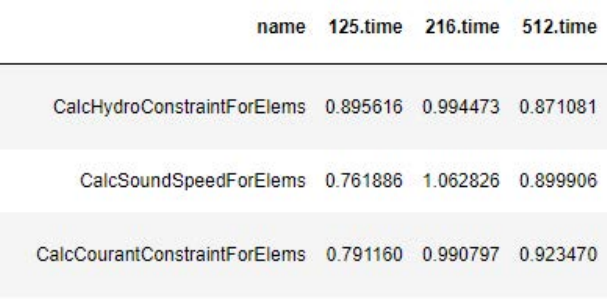}
\begin{lstlisting}[linewidth=1\columnwidth]
datasets = glob.glob("list_of_lulesh_profiles")
gfs = hatchet.GraphFrame.construct_from(datasets)
efficiency = hatchet.Chopper.speedup_efficiency(gfs, weak=True, efficiency=True)
print(efficiency.sort_values("512.time", ascending=True))
\end{lstlisting}
    \caption{Calculating efficiency of each node using the
        speedup\_efficiency function.
        The figure shows a truncated example of the returned pivot table.
        LULESH weak scaling executions running on 64, 125, 216, and
        512 processes are used for this demonstration.}
    \label{fig:spd-eff}
\end{figure}
\begin{figure*}[t]
    \centering
    \subfloat{\includegraphics[width=1\columnwidth]{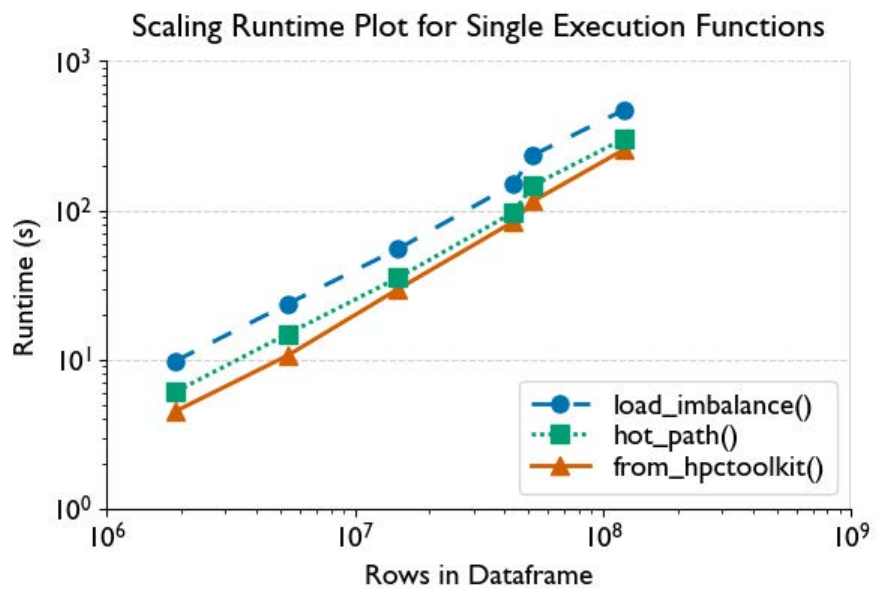}}
    \subfloat{\includegraphics[width=1\columnwidth]{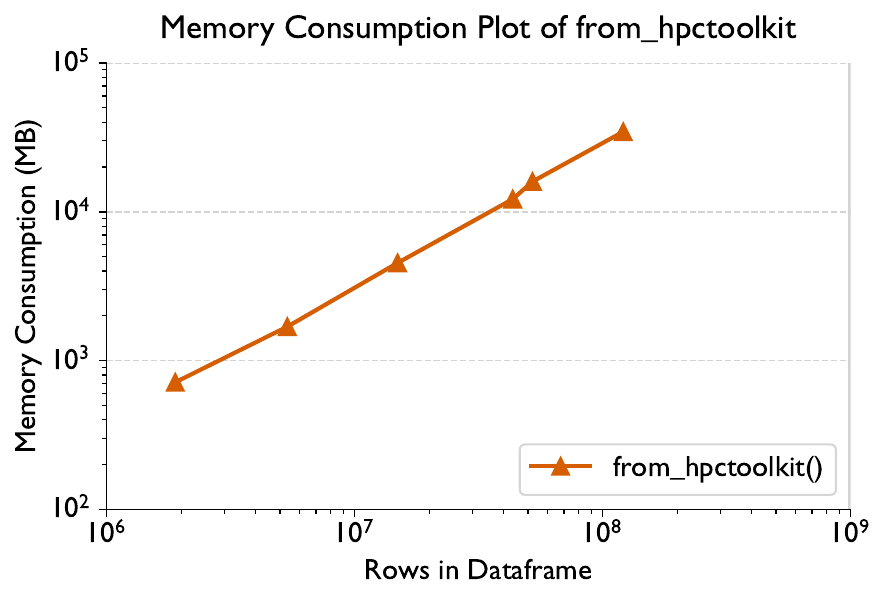}}
    \caption{Log-log plot of the runtime of the {\tt hot\_path} and
            {\tt load\_imbalance}, and {\tt
                from\_hpctoolkit} functions (left) and memory consumption of {\tt
                from\_hpctoolkit} (right). We observe that all of the functions
        scale linearly with data size and memory consumption by the
        Chopper API functions does not exceed that of file reading.}
    \label{fig:chopper_perf}
\end{figure*}
\vspace{0.08in}
\noindent{\textbf{construct\_from}}: Ingesting multiple datasets is the first
step to analyzing them. It is laborious and tedious to specify and load each profile
manually, which is necessary in Hatchet. To alleviate this
problem, we introduce {\tt construct\_from}, which takes a list of datasets
and returns a list of GraphFrames, one for each dataset. Users can then
leverage Python's built-in functionalities to create the list from names and
structures inspected from the file system.

construct\_from automatically detects the data collection source of each
profile, using file extensions, JSON schemes, and other characteristics of the
datasets that are unique to the various output formats. This allows Chopper to
choose the appropriate data read in Hatchet for each dataset, eliminating the
manual task of specifying each one. We demonstrate the power of
construct\_from in Figures~\ref{fig:construct-multirun} and~\ref{fig:spd-eff}
(line 2 in both).

\vspace{0.08in}
\noindent{\textbf{multirun\_analysis}}: Analyzing across multiple executions
typically involves comparing metric values across the individual CCT nodes of
the different executions. Implementing this manually can be cumbersome,
especially as CCTs will differ between runs. We simplify this task with the
    {\tt multirun\_analysis} function.

By default, multirun\_analysis builds a unified ``pivot table'' of the
multiple executions for a given metric. The index (or ``pivot'') is the
execution identifier. Per-execution, the metrics are also aggregated by the
function name. This allows users to quickly summarize across executions and
their composite functions for any metric.

multirun\_analysis allows flexibly setting the desired index, columns
(e.g., using file or module rather than function name), and metrics with which
to construct the pivot table. It also provides filtering of nodes below a
threshold value of the metric. The code block in
Figure~\ref{fig:construct-multirun} demonstrates multirun\_analysis with
default parameters (line 3) and its resulting table.

As we will show in Section~\ref{sec:case_study}, the multirun\_analysis
function makes it straightforward to analyze multiple executions and
significantly reduces end-user effort. Most importantly, users can easily
manipulate the pivot table programmatically or generate different ones for
different analysis tasks such as scaling and variability. In addition, it is
possible to plot the data in this pivot table with only a single line of
Python code. This is normally a laborious task to perform using only a GUI.

\vspace{0.08in}
\noindent{\textbf{unify\_multiple\_graphframes}}:

Fine-grained analysis
tasks may require preserving those individual metrics and CCT topology in
order to match them across CCT nodes.
Combining multiple large parallel profiles takes
significant programming effort. We automate this task through the
unify\_multiple\_graphframes function,
which takes multiple GraphFrames as inputs and updates each GraphFrame in place.

The unify\_multiple\_graphframes function creates a union graph object
from all input GraphFrames from the collection of unique call paths. The
updated GraphFrames point to this new object and the DataFrame of each is
updated with the missing nodes. The operation ensures that all input
GraphFrames are associated with the same unified graph and have individually
updated DataFrames.

Figure~\ref{fig:unify-multiple-graphframes} illustrates how the GraphFrames
are updated by unification.
The resulting GraphFrames share the same graph
while retaining their original metric values. Using this unified GraphFrames,
node-level (calling context-dependent) metrics can be calculated, such as
speedup and efficiency.

\vspace{0.08in}
\noindent{\textbf{speedup\_efficiency}}: Two commonly used metrics to
determine the scalability of parallel codes are {\em speedup} and {\em
        efficiency}. The {\tt speedup\_efficiency} function simplifies the task of
calculating these metrics per CCT node across multiple executions with different process or
thread counts. Given multiple GraphFrames as input, speedup\_efficiency creates a
new DataFrame with efficiency or speedup per CCT node, using
unify\_multiple\_graphframes to unify the set of nodes. An optional parameter
allows users to set a metric threshold with which to exclude unnecessary nodes.

Speedup and efficiency have different expressions under the assumption of weak
or strong scaling. Thus, the speedup\_efficiency functions should be
supplied with the type of experiment performed (weak or strong scaling) and
the metric of interest (speedup or efficiency).
Equations~\ref{eqn:speedup},~\ref{eqn:efficiency_strong}, and
~\ref{eqn:efficiency_weak} define these metrics, where $t_s$ is the
baseline value of the metric of interest, typically time spent in the
execution. In other words, $t_s$ is the metric's value in the execution with
the smallest number of processes, $s$.
$t_n$ is then the metric value from the executing with $n$ processes, where $n>s$.
Speedup for strong scaling,
$S_{\mathrm{strong}}$ is defined as the ratio of $t_s$ to $t_n$
(Eq.~\ref{eqn:speedup}) and efficiency for strong scaling, $E_{\mathrm{strong}}$ is defined by the
multiplication of the ratio of $s$ to $n$ and ratio of $t_s$ to $t_n$
(Eq.~\ref{eqn:efficiency_strong}).

\begin{equation}
    \label{eqn:speedup}
    S_{\mathrm{strong}}=\frac{t_s}{t_n}
\end{equation}
\begin{equation}
    \label{eqn:efficiency_strong}
    E_{\mathrm{strong}}=\frac{s \cdot t_s}{n \cdot t_n}
\end{equation}

We calculate only efficiency for weak scaling experiments,
$E_{weak}$, which is defined as the ratio of $t_s$ to $t_n$
(Eq.~\ref{eqn:efficiency_weak}).
\begin{equation}
    \label{eqn:efficiency_weak}
    E_{\mathrm{weak}}=\frac{t_s}{t_n}
\end{equation}

Figure~\ref{fig:spd-eff} shows the output DataFrame of efficiency values from
a weak scaling (64 to 512 process) experiment of LULESH along with the
corresponding code block (line 3). The DataFrame can then be used directly to
plot the results.

%% file: setup.tex
We collected our experiment profiles on a supercomputer
with an x86\_64 architecture and 36 cores per node.
On this machine we collected performance profiles from
LULESH 2.0.3~\cite{LULESH2:changes} and
Quicksilver 1.0~\cite{richards2017quicksilver} executions on 64 and 128
processes using 32 cores per node (2 to 16 nodes). LULESH is a proxy application
that solves a Sedov blast problem and Quicksilver solves a simplified monte
carlo problem. In addition, we strong-scaled
Tortuga using 32, 64, 128, and 256 processes.
Tortuga is a computational fluid dynamics applications provided to us
by our collaborators.
We use a
set of data collected from AMG 1.2~\cite{amg-proxy-app} (a parallel
algebraic multigrid solver) and Laghos
for a study on variability in ~\cite{nichols:ipdps2022}.
This data was collected on the same machine and
for several applications run on 512 processes with the same configuration
for almost a year. We used a subset of
the data that includes six months of HPCToolkit profiles
for AMG and Laghos executions.

We built each tool and application
using GCC 8.3.1 and Open MPI 3.0.1.
We used Score-P 7.1 to profile Tortuga and all other
applications was profiled using HPCToolkit 2021.05.15.

%% file: performance.tex
In this section, we evaluate the performance of some of
the single execution analysis functions in Chopper.

\subsection{API Performance for Single Executions}

We measure the runtime of the functions by using them with a set
of HPCToolkit profiles. The smallest GraphFrame is created
from an AMG execution on 64 processes (2 nodes) and
contains 1,893,504 rows in the DataFrame and 29,586 CCT
nodes in the graph object. The biggest is created from a
MILC execution on 256 processes (8 nodes) and contains
121,177,088 rows in the DataFrame and 473,348 CCT nodes
in the graph object. The data points in between are from
AMG executions on 128, 256, and 512 and a MILC execution
on 128 processes.

Figure~\ref{fig:chopper_perf} (a) shows the runtime of {\tt load\_imbalance}
and {\tt hot\_path} for each data size.
The runtime
of the HPCToolkit reader function, Hatchet's {\tt from\_hpctoolkit},
is included to illustrate the total time required as reading into Hatchet is
necessary to use Chopper.
The results demonstrate that all the functions
work efficiently for large profiles in terms of number of rows
in the DataFrame. The slowest function,
{\tt load\_imbalance}, takes 9.81 seconds for the
smallest and 470.94 seconds for the largest data size.
This increase in running time is expected due to the significant
increase in size of the profiles. Figure~\ref{fig:chopper_perf} (b) shows
only the memory consumption of the {\tt from\_hpctoolkit} function because
the Chopper functions consume less memory than the file reading.

%% file: case_study.tex
We demonstrate the usability and flexibility
of the analysis functionalities we provide in Chopper by analyzing
profiles from single and multiple executions.

\begin{figure}[h]
    \centering
    \includegraphics[width=1\columnwidth]{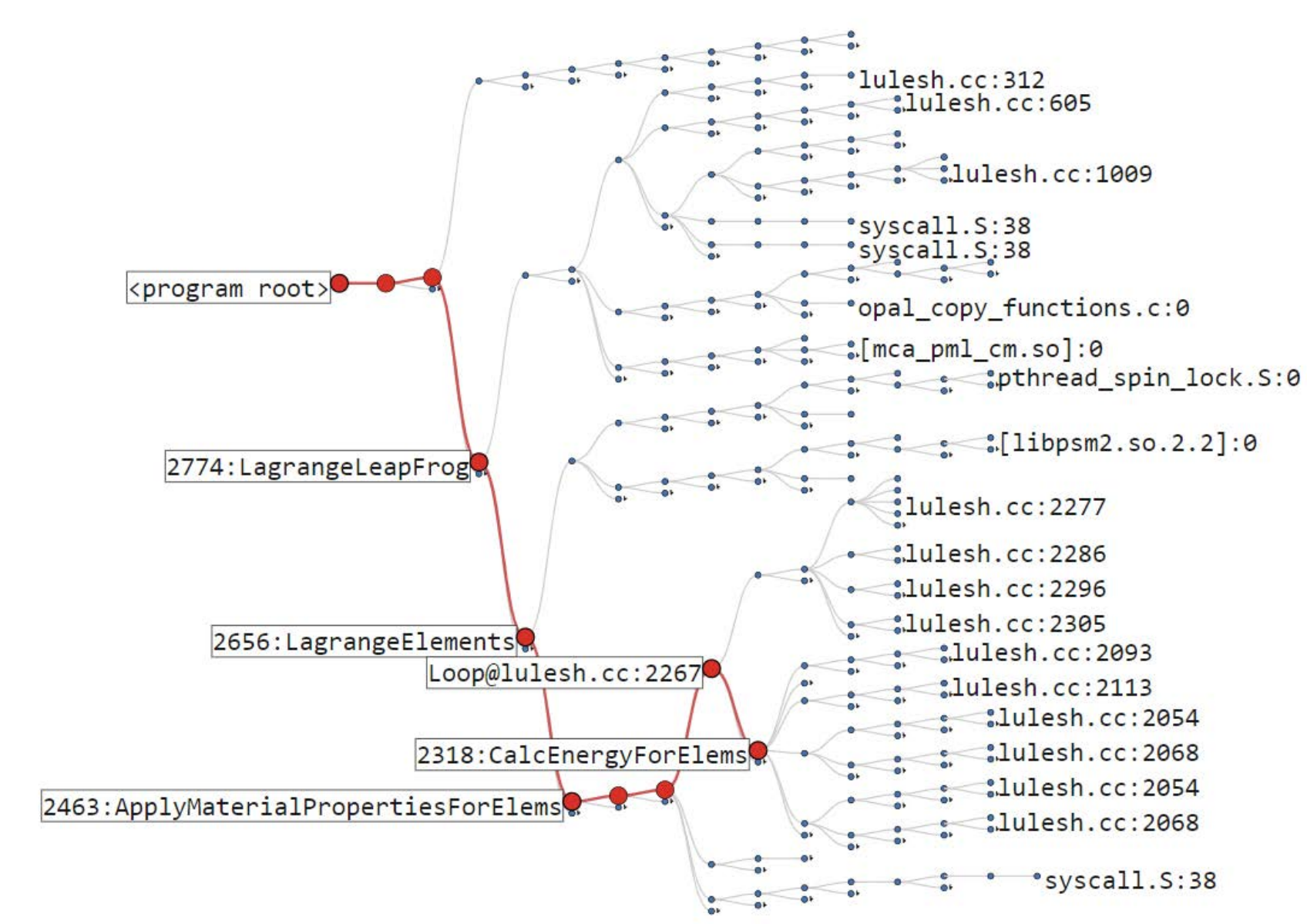}
\begin{lstlisting}[linewidth=1\columnwidth]
%load_ext hatchet.vis.loader
gf = hatchet.GraphFrame.from_hpctoolkit("lulesh_64")
hot_path = gf.hot_path()
%cct gf # Jupyter Visualization
\end{lstlisting}
    \caption{Demonstration of identifying the hot path on a
        filtered CCT gathered from a LULESH execution
        on 64 processes  using
        the {\tt hot\_path} function. We
        visualize the tree with the highlighted hot path (red coloring)
        by using Hatchet's interactive Jupyter visualization.}
    \label{fig:hotpath_case}
\end{figure}

\subsection{Analyzing a Single Execution}
Analyzing the performance of a single execution is a common task
in performance analysis. For example, the user may want to identify
the causes of performance degradation that occur when running the
application with a specific number of processes on a particular platform.
\begin{figure*}[t]
    \centering
    \subfloat[Quicksilver Load Imbalance DataFrame Output]{\includegraphics[width=1.1\columnwidth]{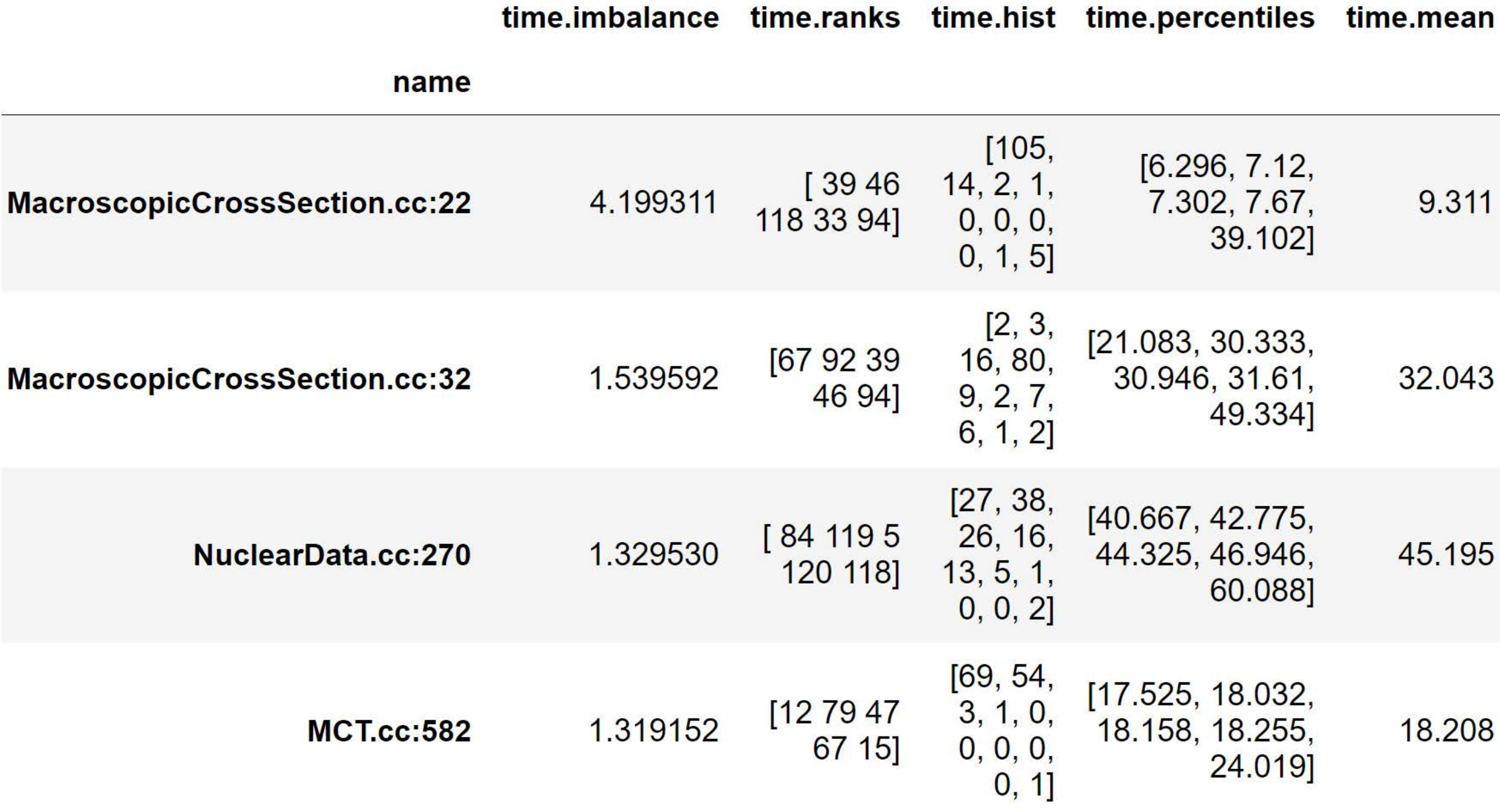}}
    \vspace{0.2in}
    \subfloat[Load Imbalance Histogram of MacroscopicCrossSection.cc:22]{\includegraphics[width=0.88\columnwidth]{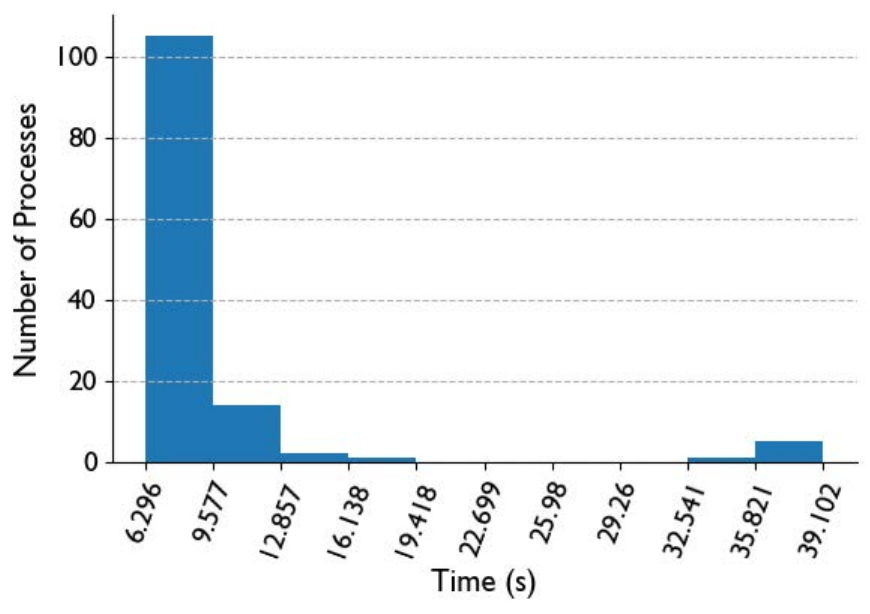}}
\begin{lstlisting}[linewidth=2\columnwidth]
graphframe = hatchet.GraphFrame.from_hpctoolkit("qs_profile_128")
graphframe_imbalance = graphframe.load_imbalance(verbose=True)
# sort the top 50 nodes that have the highest mean value by imbalance
df_imb = graphframe_imbalance.dataframe.head(50).sort_values("time.imbalance", ascending=False)
print(df_imb.head(4))  # Dataframe Output (a)
\end{lstlisting}
    \caption{Demonstration of load imbalance analysis and the results of the
    case study. The most imbalance is caused by {\tt MacroscopicCrossSection:22}.
    Chopper's {\tt load\_imbalance} function provides detailed statistics about
    the imbalance (a) that can be easily plotted by using Python libraries (b).
    We use Quicksilver execution on 128 processes.}
    \label{fig:qs_load_strong_64_256}
\end{figure*}

\begin{figure*}[t]
    \centering
    \subfloat{\includegraphics[width=1\columnwidth]{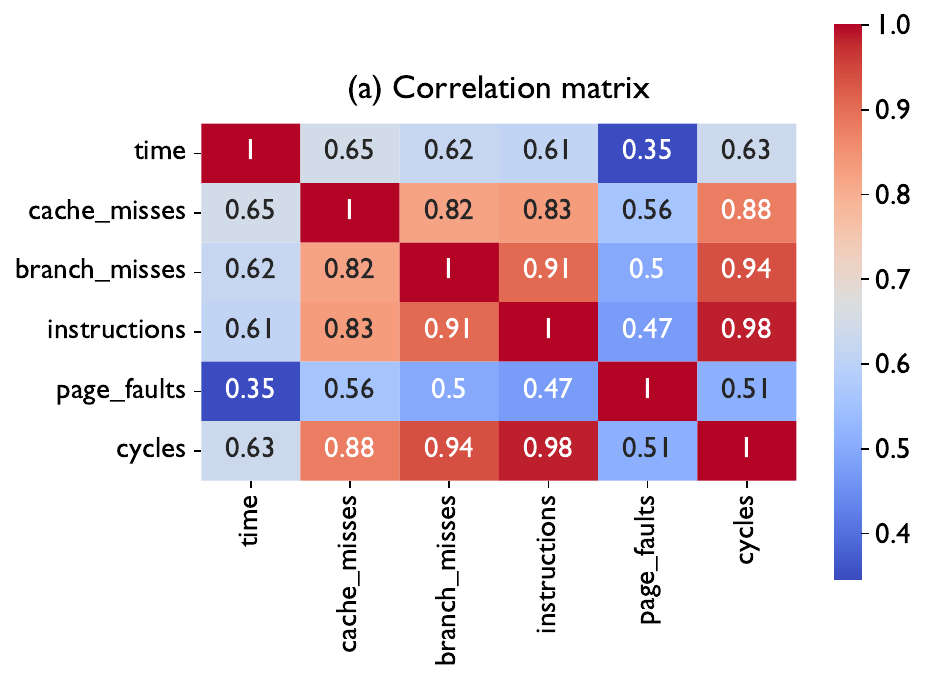}}
    \subfloat{\includegraphics[width=1\columnwidth]{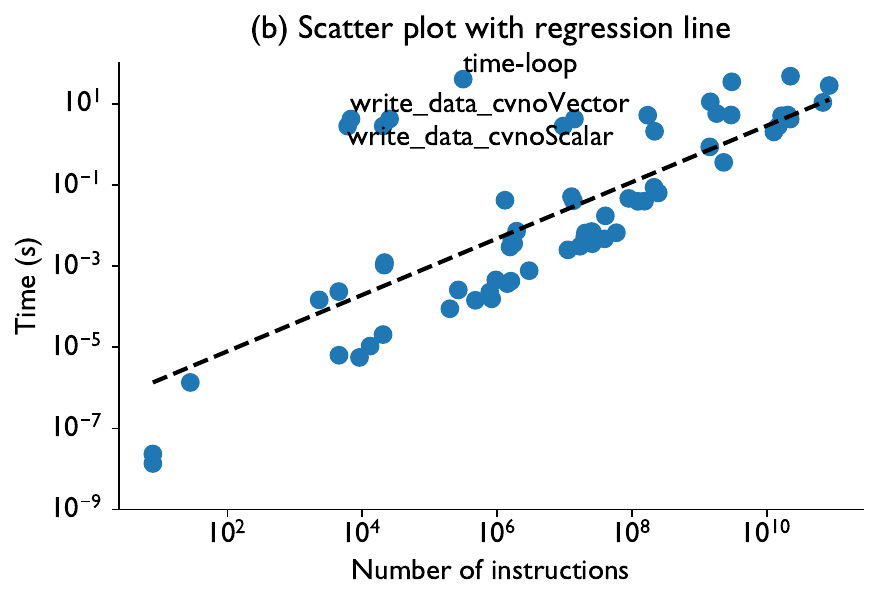}}
\begin{lstlisting}[linewidth=2\columnwidth]
correlation_matrix = gf.correlation_analysis(metrics=["time", "cache_misses", "branch_misses", "instructions", "page_faults", "cycles"], method="spearman")
sns.heatmap(correlation_matrix, annot=True, cmap='coolwarm')
gf_corr = gf.pairwise_correlation(metric1="instructions", metric2="time", logscale=True)
plt.plot(gf_corr.dataframe["instructions"], gf_corr.dataframe["time"], 'o')
plt.plot(gf_corr.dataframe["instructions"], np.exp(gf_corr.dataframe["predicted"]), '--k')
\end{lstlisting}
    \caption{Demonstration of correlation analysis performed by using Chopper. We first
        calculate the correlation matrix of all the performance matrix (a). Then, we investigate
        the relationship of instructions and time metrics at the individual CCT node level (b).
        We use the Tortuga execution on 1024 processes.}
    \label{fig:correlation}
\end{figure*}

\vspace{0.08in}
\noindent{\textbf{Finding the hot path}}:
Hot path analysis helps explore the most time-consuming call
path in the program. It may help to pinpoint potential bottlenecks.
Figure~\ref{fig:hotpath_case} shows the hot path we found in
a LULESH execution on 64 processes. We call the
hot\_path function and find the hot path. Chopper
identified \texttt{CalcEnergyForElems} as the hot node,
indicating that each of the nodes between the root and \texttt{CalcEnergyForElems}
account for 50\% or more of the inclusive time of their parents. Further exploration
can be done by examining the children and parent of the hot
node.

To visualize the hot path, we added capability to highlight the hot
path in the Jupyter visualization, which simplifies analyzing and presenting
the hot path. The visualization highlights the hot path by coloring the
nodes and edges in red and making the nodes bigger and edges thicker
than normal (Figure~\ref{fig:hotpath_case}).
The user can easily visualize the tree (line 4)
and manipulate it interactively (e.g., selecting nodes, expanding or
collapsing subtrees) for further examination and export tree state back to
Python via query.
The code block in Figure~\ref{fig:hotpath_case} demonstrates that Chopper
makes this analysis effortless with a few lines of Python code and
enables further investigation.

\vspace{0.08in}
\noindent{\textbf{Detecting load imbalance}}:
In this case study we use Quicksilver proxy application to analyze load
imbalance in an execution on 128 processes.  The code block in
Figure~\ref{fig:qs_load_strong_64_256} demonstrates how the
load imbalance analysis can be performed using Chopper.
We create the GraphFrame (line 1) and call the load\_imbalance
function with the time metric and verbose parameters (line 2). The
DataFrame associated with the returning GraphFrame is sorted by
the \texttt{time.mean} column, so that we can investigate the
load imbalance of the most time-consuming CCT
nodes. Then, we create a smaller DataFrame, \texttt{df\_imb} by
filtering out the top 50 nodes and sorting them by
\texttt{time.imbalance} (line 4).
Then, we focus on the top four nodes (line 5)
that have the highest imbalance values since the rest of nodes do not have
significant imbalance.

Figure~\ref{fig:qs_load_strong_64_256} (a) shows
the resulting DataFrame. The highest imbalance value (4.199) occurs
from \texttt{MacroscopicCrossSection:22}. The five ranks (process IDs)
with the highest time value are shown
in \texttt{time.ranks}. The \texttt{time.percentiles} column shows the
0th, 25th, 50th, 75th, and 100th percentiles. Using these two columns, we
observe that rank 39 has the most load imbalance and spends 39.1 seconds
in this function. The \texttt{time.hist} column shows the number of
processes for each node in ten equal-sized bins space across the
full range of time values.
Additionally, users can easily investigate the parent of \texttt{MacroscopicCrossSection:22}
(e.g., \texttt{df\_imb.index[0].parents}) to see where it is being called.
Using this to examine the source code, we observe that
many load and store operations are performed in \texttt{macroscopicCrossSection}
and it is called in a {\tt for} loop inside of
the \texttt{CollisionEvent} function. Therefore, uneven distribution of
load/store operations across processes may be the cause of this imbalance.

\begin{figure*}[t]
    \centering
    \subfloat{\includegraphics[width=1\columnwidth]{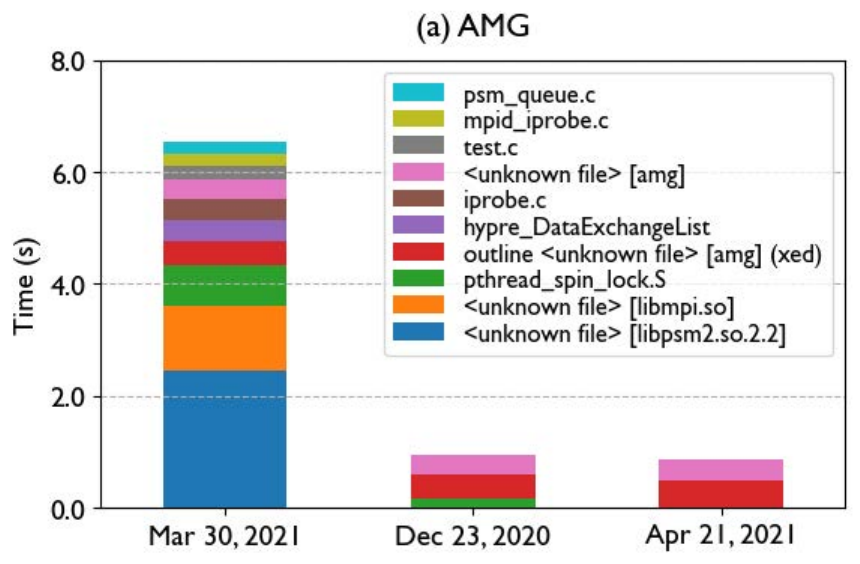}}
    \subfloat{\includegraphics[width=1\columnwidth]{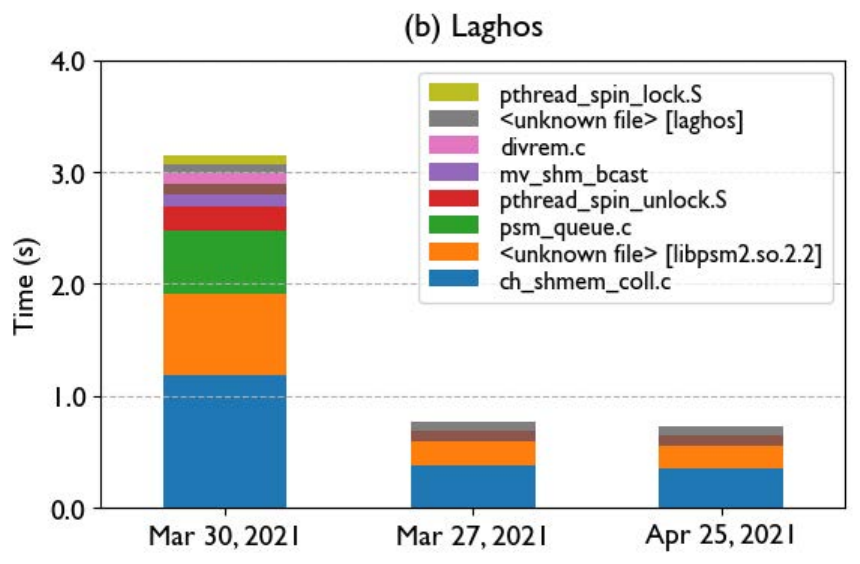}}
\begin{lstlisting}[linewidth=2\columnwidth]
datasets = glob.glob("list_of_amg_profiles")
graphframes = hatchet.GraphFrame.construct_from(datasets)
pivot_table = hatchet.Chopper.multirun_analysis(graphframes=graphframes, threshold=threshold)
pivot_table.loc[:, :].plot.bar(stacked=True)
\end{lstlisting}
    \caption{Demonstration of variability analysis by using multiple
        executions. The figure shows the executions slowest to fastest
        from left to right. We create GraphFrames for slowest,
        average, and fastest (a) AMG and (b) Laghos runs. Then, we use
        the {\tt multirun\_analysis} function to compare CCT nodes on these
        multiple executions and easily create plots by using the output of the function.
        As shown, the variation comes from communication libraries
        in both cases. All executions use 512 processes and has the same configuration.}
    \label{fig:variability}
\end{figure*}

\begin{figure*}[t]
    \centering
    \subfloat{\includegraphics[width=1\columnwidth]{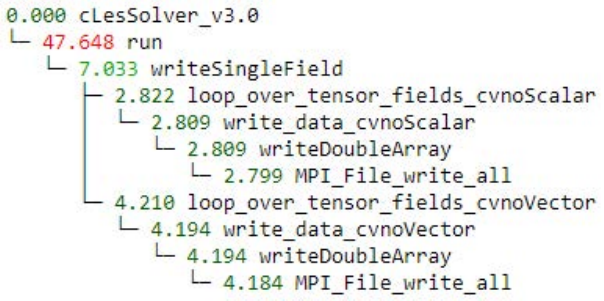}}
    \subfloat{\includegraphics[width=1\columnwidth]{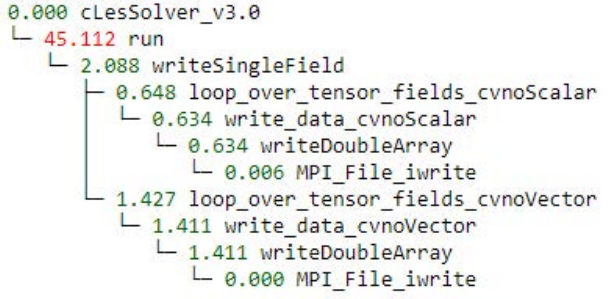}}

    \begin{minipage}[t]{0.48\textwidth}
\begin{lstlisting}[linewidth=1\columnwidth]
query = ["*", {"name": "MPI_File_write_all"}]
filtered_test = graphframe.filter(query)
print(filtered_test.tree())
\end{lstlisting}
    \end{minipage}
    \hfill
    \begin{minipage}[t]{0.49\textwidth}
\begin{lstlisting}[linewidth=1\columnwidth]
query = ["*", {"name": "MPI_File_iwrite"}]
filtered_test = graphframe.filter(query)
print(filtered_test.tree())
\end{lstlisting}
    \end{minipage}

    \caption{Call paths of the problematic portions of the program before (left)
        and after (right) the optimization. The time spent in {\tt writeSingleField}
        reduced from 7.033 to 2.088. The 1024 process count execution is used.}
    \label{fig:eff_trees}
\end{figure*}

As shown in Figure~\ref{fig:qs_load_strong_64_256} (b), most
processes spend between 6.296 to 9.577 seconds, while only a few
spend more than 32.541 seconds. This case study shows that a
detailed analysis of load imbalance at per node-granularity can be
trivially accomplished using Chopper.

\vspace{0.08in}
\noindent{\textbf{Analyzing correlation between metrics and CCT nodes}}:
In this case study, we analyze the relationship between performance metrics
and CCT nodes. We use Tortuga execution on processes. We manually annotated
and profiled Tortuga by using Score-P.

Figure~\ref{fig:correlation} (a) shows the correlation between each metric.
We first examine the correlation between performance metrics by using
the correlation\_analysis function with the Spearman method (line 1).
We create a heatmap of correlation values by using the {\tt seaborn} library in
Python (line 2). Interestingly we observe that time and other metrics are not
highly correlated. To investigate further, we use the pairwise\_correlation
function (line 4). We then create a scatter plot with a regression line by using
the output of the pairwise\_correlation function (line 5 and 6).
As shown in Figure~\ref{fig:correlation} (b), there are a few outliers CCT nodes.
The {\tt time-loop} node represents the main {\tt for} loop that includes
all the operations and functions calls on the program. The program spends relatively
significant amount of time on {\tt write\_data\_cvnoVector} and {\tt write\_data\_cvnoScalar}
although they don't execute as many instructions. Both of these functions perform parallel
IO write operations. Therefore, we observe that they perform less instructions but they
have more wait time due to IO operations. This case study demostrates that the users
can easily examine correlation between different performance metrics and investigate
outliers or potential issues by performing analyses at CCT node level. Chopper also
enables to easily plot the results by using the Python libraries.

\subsection{Comparing Multiple Executions}
More advanced analysis tasks, such as studying scalability
and variability, require analyzing multiple executions of
the same program with different parameters. In this case,
the user needs to analyze more than two CCTs.
We show that Chopper can significantly
simplify these analysis tasks.

\begin{figure}[h]
    \centering
    \includegraphics[width=\columnwidth]{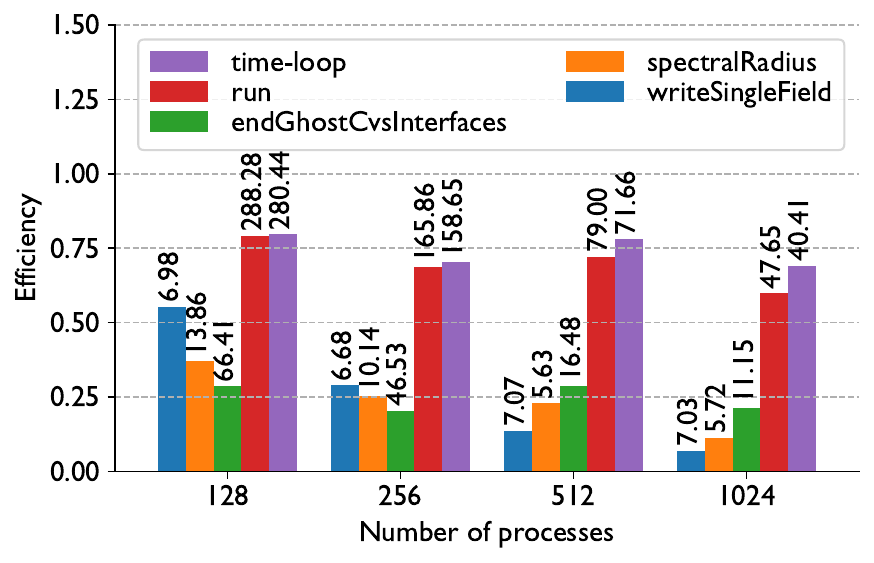}
\begin{lstlisting}[linewidth=1\columnwidth]
datasets = glob.glob("list_of_tortuga_profiles")
gfs = hatchet.GraphFrame.construct_from(datasets)
df = hatchet.Chopper.speedup_efficiency(gfs, strong=True, efficiency=True)
df = df.loc[df['1024'] < 0.7]
df.T.loc[:, :].plot.bar()
\end{lstlisting}
    \caption{Demonstration of scalability analysis by using
        multiple executions. We plot efficiency of the four least
        efficient nodes discovered in Tortuga strong scaling executions (64, 128, 256, 512, 1024 processes).
        64 process count execution is used as the baseline.
        The vertical labels on each bar corresponds the absolute time
        spent in the functions.
    }
    \label{fig:scalability}
\end{figure}

\vspace{0.08in}
\noindent{\textbf{Identifying performance variability}}:
We analyze data collected in ~\cite{nichols:ipdps2022}
that focuses on two applications,
AMG and Laghos.
The data was collected over a period of six months, during which the
applications were executed repeatedly on a fixed number of nodes using fixed
input parameters to study performance variability. In this case study, we
demonstrate how we use Chopper to quickly identify the sources of variability.
For both applications, we identify the runs that have fastest, slowest, and
average execution time and analyze profile of these runs.

Figure~\ref{fig:variability} illustrates our analysis methodology and
the resulting plots. First, we create GraphFrames
for each profile (line 2) and pass them to the
multirun\_analysis function. Using the time metric, we apply a threshold
for each of the three executions to remove the insignificant CCT nodes. Using the
table that the multirun\_analysis function constructs, we create the
plots as shown in line 4.

The resulting plots show the difference between the runs (ordered left to
right from slowest to fastest) in the execution time of the nodes causing
variability in AMG (a) and Laghos (b). These plots reveal that increase in time
on the slowest runs is caused by the communication libraries (such as
\texttt{libpsm2.s} and \texttt{libmpi.so}), which is expected due to network
congestion mentioned in the previous paper ~\cite{nichols:ipdps2022}

The Chopper API enables the analysis of multiple
executions using a single function call and presents the results in
an easy-to-plot format. This is a tedious and fraught task without
programmatic analysis capabilities as it requires comparing performance
nodes from multiple runs simultaneously. Furthermore, to the best of our
knowledge, this is the first study that uses CCT data to identify
performance variability.

\vspace{0.08in}
\noindent{\textbf{Identifying scalability bottlenecks}}:
In this case study, we analyze data from a strong scaling
experiment using Tortuga executions on 64, 128, 256, 512, and 1024 processes.
The executions use 2, 4, 8, and 16 full nodes on the supercomputer,
respectively. The efficiency at 128, 256, 512, and 1024 process counts
is calculated relative to the baseline, which is the execution with
64 processes. We used the code that we manually annotated using Score-P.

Figure~\ref{fig:scalability} demonstrates how a per-node (CCT node)
scalability analysis can be done using the Chopper API.
We first create a GraphFrame for each execution (line 2), and then call
the speedup\_efficiency function by passing all of the GraphFrames,
the metric that we want to calculate efficiency on (time by default),
the type of experiment (strong=True), and analysis type (efficiency=True)
in line 3. This function automatically unifies all the given GraphFrames
with the unify\_multiple\_graphframes function and calculates
efficiency relative to the baseline execution. We filter out the
CCT whose efficiency values are greater than 0.7 (line 4) and plot the
results by using the resulting DataFrame (line 5). In addition to
efficiency values and node names, the user can access the
corresponding file and line number from the DataFrame.

The efficiency plot (Figure~\ref{fig:scalability}) shows the nodes
that use more than 10\% of the total execution time and have
efficiency values lower than 0.7. \texttt{endGhostCvsInterfaces}
perform the communication of ghost cells in the program. Therefore,
the decreasing efficiency on these nodes indicates inefficient
communication. \texttt{spectralRadius} is called in every iteration
of the main for loop of the program. It calculates spectral radius of
a 3-dimensional tensor and calls both \texttt{MPI\_Reduce} and \texttt{MPI\_Bcast}.
\texttt{run} is a large function (772 lines of code) that includes
the main \texttt{for} loop and many IO operations. \texttt{time-loop} represents
the main for loop. \texttt{writeSingleField} includes file write operations
using all the MPI processes used.

After getting this efficiency results, we decided to focus on
the \texttt{writeSingleField} function because it is one of the functions
that has significantly decreasing efficiency. We further annotated this function
to identify the code block that cause this scalability issue. We identified
the \texttt{MPI\_File\_write\_all} function as a cause of this problem. It is a
collective and blocking function that uses all the processes on the program to
write to a file. Instead of using this collective and blocking function, we
used the nonblocking \texttt{MPI\_File\_iwrite} function and leveraged asynchrony
to optimize the function. Figure~\ref{fig:eff_trees} demonstrates
the unoptimized (left) and the optimized (right) version of the corresponding
call path. The time spent on \texttt{writeSingleField} reduced
from 7.033 to 2.088 on 1024 processes. The figure also demonstrates how to
easily get the corresponding call paths by using Hatchet's query language.

This study shows that Chopper significantly simplifies this
scalability analysis at per-node granularity by providing
functions that can automatically unify the profile outputs
and calculate efficiency. It also enables easy
plotting of the results via Python libraries.

%% file: conclusion.tex
In this study, we proposed Chopper, a Python-based API
for performance analysis, which provides programmatic
analysis capabilities that simplifies the performance analysis of
single and multiple executions of parallel applications.

We decided
to build it on top of Hatchet to leverage its programmatic
interface and visualization capabilities. We designed
the API in a way that it does not have a steep learning curve so
the users can quickly perform their analyses.

In this paper, we used several case studies to demonstrate
how Chopper enables performing common but laborious analysis
tasks by writing only a few lines of Python code. Specifically,
we presented how Chopper simplifies analysis tasks for single and
multiple executions such as detecting load imbalance, finding
hot paths, identifying scaling bottlenecks, finding correlation
between metrics and CCT nodes, and causes of
performance variation. We also demonstrated some useful
functionalities such as reading multiple profile data at
once and unifying multiple GraphFrames. We identified potential
performance problems in Tortuga and Quicksilver
applications. Additionally, we identified the performance
variability problem in AMG and Laghos runs.
The effective capabilities that Chopper provides
makes the performance analysis tasks easier to perform and
significantly reduces the effort.

In the future, we plan to improve correlation analysis by adding
predictive modeling capabilities to facilitate performance analysis.
To further simplify the analyses and reduce the effort, we plan to support
customizable plotting capabilities. Additionally, we will add support for
analyzing the performance of GPU applications.